\documentstyle[twocolumn,eqsecnum,aps]{revtex}

\begin{document}
\draft
\preprint{Draft 7}
\title{An Exactly Soluble Hierarchical
Clustering Model: Inverse Cascades,
Self-Similarity, and Scaling}
\author{A.~Gabrielov}
\address{Departments of Mathematics and Earth \&
Atmospheric Sciences\\ Purdue University, West
Lafayette, IN 47907}
\author{W.I.~Newman}
\address{Departments of Earth \& Space Sciences,
Physics \& Astronomy, and Mathematics\\
University of California, Los Angeles, CA 90095}
\author{D.L.~Turcotte}
\address{ Department of Geological
Sciences\\Cornell University, Ithaca, NY 14853 }
\date{\today}
\maketitle
\begin{abstract}
 We show how clustering as a general
hierarchical dynamical process proceeds via a
sequence of inverse cascades to produce
self-similar scaling, as an intermediate
asymptotic, which then truncates at the largest
spatial scales.
 We show how this model can provide a general
explanation for the behavior of several  models
that has been described as ``self-organized
critical,'' including  forest-fire, sandpile,
and slider-block models.
\end{abstract}
\pacs{05.65.+b, 45.70.Qj, 45.70.-n, 05.10.Cc,
05.10.-a}

\narrowtext

\input epsf.tex

\section{Introduction}
\label{sec:sec1}

 Clustering and aggregation play an important
role in many complex systems.
 In this paper, we present an inverse cascade
model for the self-similar growth of clusters.
 Elements are introduced at the smallest scale,
which then coalesce to form larger and larger
clusters.
 The inverse cascade is terminated by the loss
of the largest clusters.
 The system is thus in a quasi-steady state with
the loss of elements in large clusters balanced
by the introduction of new elements.
 The clustering process is recognized to be a
branching network similar to a DLA cluster or a
river network.
 Individual clusters are analogous to branches,
and coalescence is equivalent to the joining of
two branches.

 There is a wide range of applications for this
analysis.
 As a specific example, we consider the
forest-fire model\cite{BCT92}
 which has been said to exhibit self-organized
criticality\cite{BTW88}.
 In one version of the forest-fire model, a
square grid of sites is considered.
 At each time step, a model tree or a model
spark is dropped on a randomly chosen site.
 If the site is unoccupied when a tree is
dropped, it is ``planted.''
 The sparking frequency $f$ is the inverse
number of attempted tree drops before a spark is
dropped.
 If the spark is dropped on an empty site,
nothing happens; if it is dropped on a tree, it
ignites and ``burns'' all adjacent trees in a
model forest fire.
 In this model, individual trees are introduced
at the smallest scale, clusters of trees
coalesce to form larger and larger clusters.
 Significant numbers of trees are lost only in
the largest fires that terminate the inverse
cascade\cite{TMMN99}.
 The noncumulative frequency-area distribution
for the fires is well approximated by a
power-law relation
\begin{equation} N \propto
\frac{1}{A^\alpha}
\end{equation} with $\alpha \approx 1$.
 If the sparking frequency $f$ is relatively
large, the largest fires are relatively small
and the self-similar inverse cascade is valid
only over a relatively small range of cluster
sizes.
 If the sparking frequency $f$ is small, the
fires that terminate the cascade are large and
if $f$ is sufficiently small the fires will span
the entire grid.
 The noncumulative frequency-area distribution
of cluster sizes satisfies equation (1.1) with
$\alpha \approx 2$ and the cumulative
distribution of clusters with area larger than
$A$ satisfies equation (1.1) with $\alpha
\approx 1$.
 The behavior of the one-dimensional forest-fire
model has been discussed in terms of a cascade
by Paczuski and Bak\cite{Bak93}.
 The inverse cascade analysis is also applicable
to the sandpile model\cite{BTW88} and the
slider-block model\cite{CL89}.
 In the sandpile model the clusters are the
metastable regions that participate in
avalanches once they are triggered.
 In the slider-block model, the clusters are the
metastable regions that participate in slip
events once they are initiated.

 One of the most striking patterns in biology is
clusters or aggregations of animals\cite{PE99}.
 Examples range from bacteria to whales and
include insects, fish, and birds.
 Bonabeau et al.\cite{BDF99} showed that the
frequency-number distribution of whales satisfy
equation (1.1) with $\alpha \approx 1$.
 The model we present here should also be
applicable to these biological problems.

\section{Hierarchical Clustering}
\label{sec:sec2}

 We consider a system of stationary entities
that we shall refer to as elements.
 In terms of the forest-fire model, the elements
are the trees that are planted on a lattice.
  The system is growing due to the steady
injection of new elements that are added to
locations that are not already occupied by
previously injected elements. 
 We define connected sets of elements, i.e.\ 
groups of elements that are in contact, to be
clusters.
  Note, however, that our model does {\it not}\/
require that elements be confined to lattice
points.
  Neighbors can be defined with any metric
(e.g. distance) condition, or according to
a defined graph structure (e.g. lattice).
 In the forest-fire model, clusters are the
groups of adjacent trees that would burn in a
fire if a spark dropped on one of the trees in
the cluster.
 We construct rules for assigning rank to
clusters in such a system, based in spirit on
the Strahler\cite{S57} classification that was
originally developed for branching in river
networks.
 In this classification system, a stream with no
upstream tributaries is defined to be of rank
one; when two rank-one streams combine, they
form a stream of rank two, and so forth.
 However, when streams of different rank
combine, the rank of the dominant stream
prevails.
 Our model for the growth of clusters is an
extension of a scheme developed
earlier\cite{NNN} which only allowed for the
coalescence of clusters of the same rank.
 The new model is much richer in that it
accomodates the coalescence of clusters of all
ranks and can, therefore, describe a much wider
array of phenomena.

 The rules for our cluster model are:
\begin{enumerate}
\item We define a single element that is added
to a system to be a cluster of rank 1.
\item If a new element is added adjacent to an
existing cluster, we say that it is added to the
cluster without changing that cluster's rank,
unless the cluster is a single element.
 In that special case, we define the two
elements as forming a cluster of rank 2.
\item If a new element connects two existing
clusters of ranks $i$ and $j$, respectively,
then the rank of this new cluster is defined as
$i+1$ when
$i=j$ and as
$\max\left\lbrace i,j\right\rbrace$ when
$i\ne j$.
 In words, this is equivalent to saying that
when two clusters of equal rank coalesce, then
the rank increases by one; however, if the two
clusters are not of equal rank, then the rank of
the larger cluster prevails.
\item If a new element connects three or more
clusters, then the rank of the new cluster is
defined to be
\begin{itemize}
\item the maximal rank of these clusters, when
one of the clusters has a rank exceeding that of
all of the others, or
\item the maximal rank of these clusters plus
one, when there are two or more clusters of the
same maximal rank.
\end{itemize}
\indent[This is a rare event---akin to a
four-body interaction---and it is ignored in the
model equations given below.]
\item We terminate the inverse cascade of
elements from small to large clusters by
eliminating clusters of a specified high rank.
\end{enumerate}
\noindent In Figure 1, we illustrate  how this
model works.

 We now wish to establish the dynamical
equations governing the evolution of this system.
 Let us define $N_i$ to be the number of
clusters with rank
$i$, for $i\ge 1$.
 Let $m_i$ be the average mass---i.e., the
number of elements---of a cluster of rank
$i$.
 Then, the total mass $M_i $ of the clusters of
rank $i$ is given by
\begin{equation} M_i=N_i m_i\quad.
\end{equation}
 For convenience, we will define the mass of a
single element to be one, namely
$m_1=1$.
 For example, in two dimensions, we can regard
$m_i$ as the mean area $A_i$ of a cluster of rank
$i$.
 This would be the case in the forest-fire model.

 We now develop a mean-field approximation
describing the dynamical evolution prescribed by
the mapping rules given above.
 As indicated, we ignore the simultaneous
coalescence of more than two clusters.
 We denote the instantaneous change in all
quantities using the mapping symbol
$\mapsto$.
 Accordingly, when two clusters of ranks $i$ and
$j$ coalesce, the values of
$N_i$ and $M_i$ are modified as follows.
 For $i=j$,
\begin{equation}N_{i+1}\mapsto N_{i+1}+1,\quad
N_i\mapsto N_i-2,
\end{equation}
\begin{equation} M_{i+1}\mapsto M_{i+1}+2
m_i,\quad M_i\mapsto M_i-2 m_i,
\end{equation} and for $i<j$,
\begin{equation} N_i\mapsto N_i-1,\quad
N_j\mapsto N_j,
\end{equation}
\begin{equation} M_j\mapsto M_j+m_i,\quad
M_i\mapsto M_i-m_i,
\end{equation} with equivalent expressions for
$j<i$.
 In these equations for
$M_j$, we have ignored the addition of an
element that bridges or joins the two clusters.
 Since $m_i$ will be shown to increase in an
essentially geometric progression with respect
to the rank $i$, the omission of that solitary
unit mass in the calculation does not influence
the asymptotic properties as $i \rightarrow
\infty$.

 In our model, coalescence occurs when a new
element connects two existing clusters.
 (We have already indicated that 4-body and
higher order effects will be neglected.)
 Accordingly, in the mean field approximation,
we assume that the rate
$r_{ij}$ of coalescence between clusters of ranks
$i$ and $j$ is proportional to the product of
their total numbers, $N_i$ and $N_j$, and to the
product of their boundary sizes,
$\ell_i$ and
$\ell_j$, and is naturally related to the joint
probability of the new element connecting two
pre-existing clusters.
 For example in two dimensions, $\ell_i$ refers
to the effective length of the cluster boundary.
 Thus, we assume that
\begin{equation} r_{ij} \propto N_i
\ell_i N_j \ell_j\quad.
\end{equation}
 This is an Euclidean approximation, and emerges
in the spirit of classical kinetic theory,
although the mechanics of this problem is
entirely different.
 In sections IV and VII, this model will be
modified to accommodate the possible fractal
geometry of clusters.

 We now define
\begin{equation} L_i=N_i \ell_i
\end{equation}
 to be the total size of the boundary associated
with clusters of rank $i$.
 We select the normalization for our time-scale
so that
$r_{ij}=L_i L_j$.
 Accordingly, let $C$ be the injection rate of
single elements, utilizing this time scale.
 The evolution of the system can be determined
by appropriately adapting equations (2.2)--(2.5).
 From equations (2.2) and (2.4), we write
\begin{equation}
\dot N_1=C-2 L_1^2-\sum_{j=2}^\infty L_1 L_j,
\end{equation}
\begin{equation}\dot N_i=L_{i-1}^2-2
L_i^2-\sum_{j=i+1}^\infty L_i L_j,\quad{\rm
for}\ i>1.
\end{equation}
 In equation (2.8), we observe that the rate of
change in the number of clusters of rank 1 is
equal to the injection rate minus the rate of
coalescence of rank 1 clusters together with the
rate of coalescence of rank 1 clusters with
clusters of larger rank.
 The factor of 2 appears because {\it two}\/
rank 1 clusters were lost in coalescing to form
a rank 2 cluster.
 Meanwhile, in equation (2.9), we observe that
the rate of change in the number of clusters of
rank $i$ is equal to the rate of rank $i$
cluster formation from the coalescence of pairs
of rank $i-1$ clusters, minus the rate of
coalescence of pairs of rank $i$ clusters,
together with the rate of coalescence of rank $i$
clusters with clusters of larger rank $j>i$.

 In a similar way, taking into account
$m_1=1$, we can express the mass-balance in the
system, derived from equations (2.3) and (2.5),
according to
\begin{equation}
\dot M_1=C-2 L_1^2-\sum_{j=2}^\infty L_1 L_j,
\end{equation}
\begin{eqnarray}
\dot M_i=2 L_{i-1}^2
m_{i-1}&+&\sum_{k=1}^{i-1}L_i L_k m_k -2 L_i^2
m_i\nonumber\\&-&\sum_{j=i+1}^\infty L_i L_j
m_i,\quad{\rm for}\ i>1.
\end{eqnarray}
 Note that equations (2.8) and (2.10) are
identical since $M_1=N_1$.

 We observe that the equations above have the 
potential for self-similarity, since most of the
sums are infinite in extent, and might be
expected to be convergent.
 Intuitively, we expect that $L_j$ will diminish
as $j$ increases; while the boundary size of
individual clusters of rank
$j$ increase, their absolute numbers will
decrease even more rapidly so that the total
boundary size in clusters of rank $j$ will be
monotone decreasing.
 The finite sum, which appears in equation
(2.11), is somewhat more involved.
 Nevertheless, it is reasonable to expect that
the product of $	m_k$ with $L_k$ will steadily
diminish as $k$ becomes smaller and that
negligible contributions emerge from low values
of $k$.
 Finally, it is easy to see that all of the
governing rate equations will quickly converge,
in the sense of an inverse cascade from
$i=1$ to some finite cut-off, as
$t\rightarrow \infty$.
 As $N_1$ begins to grow, it provides a stimulus
to the growth of $N_2$, and so on.
 Similarly, as the masses at each rank in the
system grow, they will in turn cause the
boundary size $\ell_i$ of each cluster of rank
$i$ to grow, basically in proportion to some
power in
$m_i$.
 With this intuition in hand, we now obtain the
steady-state solution for this system.

\section{Steady State Solution:  Cluster and
Mass Scaling}

 We derive a steady state solution for an
inverse cascade from equations (2.8) through
(2.11).
 In our inverse cascade, single elements are
introduced at the lowest level, and they
coalesce to form larger and larger clusters.
 The inverse cascade is terminated by assuming
that very large clusters are removed from the
system.
 We assume that our system develops in a
sufficiently large region, so that edge effects
can be ignored over a long time.
 Otherwise, we will have a completely
space-filling solution and percolation effects
will govern.
 We can regard this (limited) steady-state
solution  to be an intermediate
asymptotics\cite{ZB} for our system---our
solution will describe the similitude that
emerges before percolation and space-filling
issues become significant.
 The steady state solution follows when the time
derivatives in the left hand sides of equations
(2.8)--(2.11) vanish with the result
\begin{equation} C=2 L_1^2+\sum_{j=2}^\infty L_1
L_j.
\end{equation}
\begin{equation} L_{i-1}^2=2
L_i^2+\sum_{j=i+1}^\infty L_i L_j,\quad{\rm
for}\ i>1.
\end{equation}
\begin{eqnarray} 2 L_{i-1}^2
m_{i-1}&+&\sum_{k=1}^{i-1}L_i L_k m_k=
\nonumber\\2 L_i^2 m_i&+&\sum_{j=i+1}^\infty L_i
L_j m_i,\quad{\rm for}\ i>1.
\end{eqnarray}
 As noted earlier, equations (2.8) and (2.10)
are equivalent.

 Equation (3.2) has a self-similar solution,
since that equation is invariant under
$i\mapsto i+1$, and depends only on
$L_j/L_i$.
 Thus, we seek a solution having the form
\begin{equation} L_i=a x^{i-1}
\end{equation}
 where $0<x<1$.
 The first of these constraints on $x$
corresponds to boundary sizes being positive,
while the second is necessary for the summation
to exist.
 We find that $x$ satisfies
\begin{equation} 2 x^{2i-2}+\sum_{j=i+1}^\infty
x^{i+j-2}=x^{2i-4}.
\end{equation}
 Summing the infinite geometric series
explicitly and dividing by
$x^{2i-4}$, we obtain
\begin{equation} 2x^2+{x^3\over 1-x}=1,\quad{\rm
or}\quad x^3-2x^2-x+1=0.
\end{equation}
 This equation has a single root in the range
$0<x<1$, namely $x=0.55495813...$ . Given
equations (3.1) and (3.6), we find that
\begin{equation} C=a^2 \left[2
+x/\left(1-x\right)\right]=a^2 {\rm\quad or\quad}
a=C^{1/2}\quad.
\end{equation}
 Substitution of these results into equation
(3.4) gives
\begin{equation} L_i = C^{1/2} \left( 0.55495813
\right)^{i-1} \quad.
\end{equation}

 We now turn our attention to equation (3.3).
 We substitute equation (3.4) into equation
(3.3), dividing by $a^2 x^{i-3}$ and taking into
account equation (3.6).
 We then obtain
\begin{eqnarray} 2x^{i-1} m_{i-1}
+\sum_{k=1}^{i-1} x^{k+1} m_k&=&\nonumber\\
2x^{i+1} m_i +{x^{i+2}\over 1-x} m_i&=&x^{i-1}
m_i.
\end{eqnarray}
 This equation does not have an exactly
self-similar solution, since it is not invariant
under
$i\mapsto i+1$.
 Suppose that we make the substitution
\begin{equation} x^{i-1} m_i=y^{i-1}\quad,
\end{equation}
 assuming that $y>1$, whereupon we obtain from
summing the finite series
\begin{equation}
2xy^{i-2}+x^2\cdot{y^{i-1}-1\over y-1}=y^{i-1}.
\end{equation}
 We observe that the solution for $y$ in this
equation depends upon $i$.
 However, for large $i$, equation (3.11)
approximately implies, assuming that we can
replace $y^{i-1} -1$ by $y^{i-1}$, that
$2x+x^2 y/(y-1) = y$ which we rewrite as
\begin{equation} y^2-(x+1)^2 y+2x=0.
\end{equation}
 This equation has a unique solution for $y>1$,
namely
$y=1/x=1.8019377...$ .
 Accordingly, for large $i$, we have asymptotic
self-similarity with
\begin{equation} m_i\approx \alpha x^{1-i}
y^{i-1}=\alpha c^{i-1},
\end{equation}
 where $c=1/x^2=3.24697602...$ .
 With $m_1 = 1$, we have
\begin{equation} m_i \approx \left( 3.24697602
\right)^{i-1}\quad.
\end{equation}

 Before moving to issues dealing with fractals
and branching, the solutions we have just
obtained for $L_i$ and for $m_i$ can be
immediately exploited.
 Since $L_i \propto x^i$ and, approximately,
$m_i \propto x^{-2i}$, we observe that
$L_i\sqrt{m_i} \approx\rm const.$
 For example in two dimensions, recalling that
$L_i\equiv N_i
\ell_i$ and introducing the Euclidean relation
that
$\ell_i \propto \sqrt{m_i}$, it follows that
$N_i m_i\approx\rm const.$ or, equivalently, we
find the number-mass or number-area relationships
\begin{equation} N_i \propto 1/m_i \propto
1/A_i\quad.
\end{equation}
 This is equivalent to equation (1.1) with
$\alpha = 1$.
 The branch numbers $N_i$ are loosely equivalent
to a logarithmic binning of cluster sizes.
 Logarithmic binning is equivalent to a
cumulative distribution.
 Thus, the result given in equation (3.15)
 is in agreement with the distribution of
cluster sizes obtained from the forest-fire
model as discussed above.
 The concept of clusters can also be extended to
both sandpile and slider-block models.
 In these cases, the clusters are the metastable
regions that will avalanche or slip when an
event is triggered.
 In both cases, the cumulative distribution of
cluster sizes satisfy equation (1.1) with
$\alpha \approx 1$.
 These scaling relationships are archetypical of
self-organized criticality.
 Remarkably, this scaling has been deduced using
solely analytic means from our inverse-cascade
hierarchical cluster model.

\section{Adaptation for Fractal Perimeter:
Cluster and Mass Scaling}

 In the analysis given in the previous sections,
we assumed that the rate of cluster coalescence
$r_{ij}$ was proportional to the linear
dimensions of the two clusters as given in
Equation (2.6).
 We now generalize this dependence to account
for the possibility of fractal clusters by
introducing an ``efficiency'' factor
$\epsilon<1$, with an appropriate scaling such
that
\begin{equation}
r_{ij}\approx\epsilon^{-\left\vert
j-i\right\vert}N_i \ell_i N_j
\ell_j=\epsilon^{-\left\vert j-i\right\vert}L_i
L_j\quad.
\end{equation}
 As before, $r_{ij}$ is the rate of coalescence
between clusters of ranks $i$ and $j$.
 This modification can, for example, describe
the increased efficiency with which a smaller
cluster can coalesce with a larger one, since
the smaller cluster can become attached inside
one of the nooks and crannies that can
characterize a fractal perimeter.

 With this modification, we obtain analogs of
equations (2.8)--(2.11)
\begin{equation}
\dot N_1=\dot M_1=C-2 L_1^2-\sum_{j=2}^\infty
\epsilon^{1-j}L_1 L_j,
\end{equation}
\begin{equation}
\dot N_i=L_{i-1}^2-2 L_i^2-\sum_{j=i+1}^\infty
\epsilon^{i-j}L_i L_j,\quad{\rm for}\ i>1,
\end{equation}
\begin{eqnarray}
\dot M_i&=&2 L_{i-1}^2
m_{i-1}+\sum_{k=1}^{i-1}\epsilon^{k-i}L_i L_k
m_k\nonumber\\& -&2 L_i^2 m_i-\sum_{j=i+1}^\infty
\epsilon^{i-j}L_i L_j m_i,\quad{\rm for}\ i>1.
\end{eqnarray}
 In the steady state, we obtain analogs of
equations (3.1)--(3.3)
\begin{equation} C=2 L_1^2+\sum_{j=2}^\infty
\epsilon^{1-j}L_1 L_j.
\end{equation}
\begin{equation} L_{i-1}^2=2
L_i^2+\sum_{j=i+1}^\infty
\epsilon^{i-j}L_i L_j,\quad{\rm for}\ i>1.
\end{equation}
\begin{eqnarray} 2 L_{i-1}^2
m_{i-1}&+&\sum_{k=1}^{i-1}\epsilon^{k-i}L_i L_k
m_k=\nonumber \\ 2 L_i^2 m_i
&+&\sum_{j=i+1}^\infty
\epsilon^{i-j}L_i L_j m_i,\quad{\rm for}\ i>1.
\end{eqnarray} 
 Substituting equation (3.4) into (4.6) we
obtain an analog of (3.6)
\begin{eqnarray} 2 x^2+{x^3\over\epsilon-x}=1,
\quad&{\rm or}&\quad x^3-2\epsilon
x^2-x+\epsilon=0,\nonumber\\
\quad&{\rm or}&\quad \epsilon={x-x^3\over 1-2
x^2}\quad.
\end{eqnarray}
 As the $L_i$ are positive, $x$ must be
positive from its definition in (3.4).
 Suppose that $\epsilon=x$. 
 Then $x-2x^3=x-x^3$,
giving
$x=0$,  a contradiction. 
 Accordingly, for positive $x$, the sign
of $\epsilon-x= x^3/\left(1-2x^2\right)$ changes
only at
$x=\sqrt{1/2}$ where $\epsilon$ changes sign as
it passes through infinity (due to the
denominator).
 It is easy to see that the sign of both
$\epsilon-x$ and $\epsilon$ is positive for
$0<x<\sqrt{1/2}$, and negative for
$x>\sqrt{1/2}$.
 As $x<\epsilon$ is nesessary
for the summation of the geometric series to
exist, this implies that
$x<\sqrt{1/2}$. 
 In addition, the condition
$\epsilon<1$ requires that
$x<0.55495813\ldots$. 
 For
example, $x=0.5$ corresponds to
$\epsilon=3/4$.
 From equations (4.5) and (4.8), we obtain that
$a=x\,C^{1/2}$, for any $\epsilon$.

 Let us turn now to the mass balance equation
(4.7).
 Substituting (3.4) and assuming
$x^{i-1}m_i=y^{i-1}$, we obtain an analog of
equation (3.11)
\begin{equation} 2 x
y^{i-2}+x^2\epsilon^{1-i}{(\epsilon
y)^{i-1}-1\over\epsilon y-1} =y^{i-1}.
\end{equation}
 We assume that
$\epsilon y>1$.
 Precisely as in equation (3.11), we observe
that the solution for $y$ in this equation
depends upon
$i$.
 However, for large $i$, equation (4.9)
approximately implies that $2 x +x^2 y/(\epsilon
y-1)=y$ which we rewrite as
\begin{equation}
\epsilon y^2-(x^2+2\epsilon x+1)y+2x=0.
\end{equation}
 Due to equation (4.8),  equation (4.10) has a
solution
$y=1/x$, for any $\epsilon$. Note that condition
$\epsilon y>1$ is satisfied for $y=1/x$.
Accordingly, for large $i$, we have asymptotic
self-similarity with
$m_i\approx\alpha c^{i-1}$, where
$c=1/x^2$, as in equation (3.13).
 For example, when
$\epsilon=3/4$, we have $c=4$.

 It is important to remember that $\epsilon$
describes the perimetric fractal scaling for the
clusters.
 The relationship between perimetric and areal
scaling remains a controversial topic.
 However, assuming that one can identify an
appropriate link between the two, for example in
the context of forest fire or other models, then
the preceding discussion makes it possible to
identify the frequency-area relationship for
fractal clusters, in analogy to the $N \propto
1/A$ relationship we identified previously for
Euclidian clusters.

\section{Branching Numbers}

 In the analogy between clustering and river
networks that we have discussed above, we can
write for our clusters
\begin{equation}
\frac{N_{i+1}}{N_i} = x^2
\end{equation}
 which is known as the bifurcation ratio for
river networks.
 Also, we have
\begin{equation}
\frac{\ell_{i}}{\ell_{i+1}} = x
\end{equation} which is known as the
length-order ratio for river networks.
 For river networks, the fact that these two
ratios are almost constant is known as Horton's
laws\cite{H45}.

 A major step forward in classifying river
networks was made by Tokunaga\cite{T78}.
 He extended the Strahler ordering system to
include side branching.
 A first-order branch joining another
first-order branch is denoted by the subscript
``11'' and the number of such branches is
$N_{11}$, a first-order branch joining a
second-order branch is subscripted ``12'' and
the number of such branches is $N_{12}$; a
second-order branch joining a second-order
branch is subscripted ``22'' and the number of
such branches is $N_{22}$.

 In order to apply the concept of side branching
to the coalescence of clusters, let us suppose
that we have a coalescence of two clusters, of
ranks $i$ and $j$.
 In the case $i<j$, the cluster of rank $i$
becomes a {\it branch}\/ of the cluster of rank
$j$.
 Note that, if the smaller cluster has its own
branches, these branches are {\it not}\/ counted
as branches of the larger cluster.
 However, these branches, together with all of
their branches, etc. are counted as {\it
subclusters} of the larger cluster.
 In analogy to river networks, branches are to
tributaries as clusters are to drainage basins.
 A branch formed by the cluster of rank $i$ is
considered to be a subcluster too, and is
assigned the rank $i$.
 Any other subcluster is assigned the rank of a
cluster from which it first formed as a branch.
 In analogy to river networks, subclusters of a
cluster correspond with the streams in a
drainage basin.
 The case
$i>j$ is treated similarly.
 In the case $i=j$, both clusters of rank
$i$ become branches of rank $i$ of the new
cluster of rank $i+1$.
 Subclusters and their ranks are defined the
same way as above.

 Let $t_{ij}$ be the average number of branches
of rank $i$ in a cluster of rank
$j$, for $i<j$, and let $n_{ij}$ be the total
number of sub-clusters of rank $i$ in a cluster
of rank $j$.
 For $i=j$, we define $t_{ii}=n_{ii}=1$.
 By definition, for $i<j$ we have
\begin{equation}
n_{ij}=\sum_{k=i}^{j-1}n_{ik}t_{kj}\quad.
\end{equation}
 Moreover, let
$N_{ij}=N_j n_{ij}$ be the total number of
sub-clusters of rank $i$ for all clusters of
rank $j$, and let $T_{ij}=N_j t_{ij}$ be the
total number of branches.
 This classification scheme is illustrated  in
Figure 2.
 In (a), we have a cluster of rank ``1'' which
corresponds to a single tree in the forest-fire
model.
 In (b), two clusters of rank ``1'' have
coalesced to form a cluster of rank ``2.''
 This cluster has been joined by a cluster of
rank ``1.''
 In the forest-fire model, two trees on adjacent
grid points have been joined by a third tree.
 In (c) and (d), clusters of rank ``3'' and
``4'' are illustrated.
 For this example, we have $n_{12} = n_{23} =
n_{34} =3$, $n_{13} =n_{24}=11$, $n_{14}=43$,
$t_{12}=t_{23}=t_{34}=3$,
$t_{13}=t_{24}=2$, and $t_{14}=4$.

 As before, we regard the coalescence of more
than two clusters as being exceedingly rare and
neglect them in our treatment.
 When two clusters, of ranks $i$ and $j$
coalesce, we prescribe the mappings for
$N_{ki},\ N_{kj},\ T_{ki}$, and $T_{ij}$ as
described below.
 When $i=j$,
\begin{equation} N_{k,i+1}\mapsto N_{k,i+1}+2
n_{ki},\quad N_{ki}\mapsto N_{ki}-2 n_{ki},
\quad{\rm for}\ k<i ,
\end{equation}
\begin{equation} T_{i,i+1}\mapsto
T_{i,i+1}+2,\quad T_{k,i}\mapsto T_{ki}-2 t_{ki},
\quad{\rm for}\ k\le i ;
\end{equation} and when $i<j$,
\begin{equation} N_{kj}\mapsto
N_{kj}+n_{ki},\quad N_{ki}\mapsto N_{ki}-n_{ki} ,
\quad{\rm for}\ k\le i ,
\end{equation}
\begin{equation} T_{ij}\mapsto T_{ij}+1,\quad
T_{ki}\mapsto T_{ki}-t_{ki},\quad{\rm for}\ k\le
i.
\end{equation}

 Given the rate of coalescence $r_{ij}=L_i L_j$,
we describe the time evolution of the branching
process by the following equations
\begin{eqnarray}
\dot N_{kj}=2 L_{j-1}^2 n_{k,j-1} &+&
\sum_{i=k}^{j-1}L_i L_j n_{ki} -2 L_j^2 n_{kj}
\nonumber\\
 &-&\sum_{i=j+1}^\infty L_i L_j n_{kj},\quad{\rm
for}\ k<j,
\end{eqnarray} from equations (5.4) and (5.6),
and
\begin{eqnarray}
\dot T_{j-1,j}&=&2 L_{j-1}^2+L_{j-1} L_j-2 L_j^2
t_{j-1,j}\nonumber\\ &-&\sum_{k=j+1}^\infty L_k
L_j t_{j-1,j},
\quad{\rm for}\ j>1,
\end{eqnarray}
\begin{equation}
\dot T_{ij}=L_i L_j-2 L_j^2
t_{ij}-\sum_{k=j+1}^\infty L_k L_j
t_{ij},\quad{\rm for}\ i<j-1,
\end{equation} from equations (5.5) and (5.7).
 As before, we turn our focus to the steady
state solution of equations (5.8) through (5.10).

\section{Steady State:  Branching Numbers}

 We begin for the steady state case by setting
the time derivatives in the left hand sides of
equations (5.8)--(5.10) to zero.
 We obtain
\begin{eqnarray} 2 L_{j-1}^2
n_{k,j-1}&+&\sum_{i=k}^{j-1}L_i L_j n_{ki}
=\nonumber\\2 L_j^2 n_{kj}&+&\sum_{i=j+1}^\infty
L_i L_j n_{kj},\quad{\rm for}\ k<j.
\end{eqnarray}
\begin{eqnarray} 2 L_{j-1}^2+L_{j-1}
L_j&=&\nonumber\\2 L_j^2 t_{j-1,j}
&+&\sum_{k=j+1}^\infty L_k L_j t_{j-1,j},
\quad{\rm for}\ j>1.
\end{eqnarray}
\begin{equation} L_i L_j=2 L_j^2
t_{ij}+\sum_{k=j+1}^\infty L_k L_j
t_{ij},\quad{\rm for}\ i<j-1.
\end{equation}
 We observe that, due to the finite summation
present in equation (6.1), it is not invariant
under $j-k \mapsto j-k+1$ and its solution is
not exactly self-similar in
$j-k$.
 However, we now employ the same methodology
used in \S III and obtain asymptotically valid
approximate solution.
 In particular, we substitute (3.4) into
equation (6.1) and divide by $a^2 x^{j+k-4}$,
and we obtain
\begin{eqnarray}
2x^{j-k}n_{k,j-1}&+&\sum_{i=k}^{j-1}x^{i-k+2}n_{ki}=\nonumber\\
2x^{j-k+2}n_{kj}&+&{x^{j-k+3}\over 1-x}
n_{kj}=x^{j-k}n_{kj}.
\end{eqnarray}
 Based on our result obtained using equation
(3.10), we introduce
\begin{equation}x^{j-k} n_{kj}=z^{j-k}
\end{equation} assuming $z>1$, and we obtain
from summing the finite series in equation (6.4)
\begin{equation}
2xz^{j-k-1}+x^2\cdot{z^{j-k}-1\over z-1}=z^{j-k}.
\end{equation}
 Approximating $z^{j-k} -1$ by
$z^{j-k}$ in the asymptotic limit $j\gg k$,
equation (6.6) approximately implies that
$2x+x^2 z/(z-1)=z$, or
\begin{equation} z^2-(x+1)^2 z+2x=0.
\end{equation}
 This latter equation is {\it identical}\/ to
equation (3.12), and has a unique solution
$z>1$, namely
$z=1/x=1.8019377... $ and, thereby, demonstrates
that the branching network description preserves
the same structural character.
 Accordingly, for
$j\gg k$, we have
\begin{equation} n_{kj}\approx\beta x^{k-j}
z^{j-k}=\beta c^{j-k},
\end{equation}
 where
$c=1/x^2=3.24697602...$ as before.
 Thus, we have approximately
\begin{equation} n_{kj} \approx \left(
3.24697602 \right)^{j-k}
\end{equation} in the limit $j \gg k$.
 For the deterministic example given in Fig.\ 1,
we have $n_{kj} \approx 4^{j-1}$ for $j \gg k$.
 Substituting (3.4) into equation (6.3) and
dividing by $a^2 x^{2j-4}$, we obtain
\begin{equation} x^{i-j+2}=2x^2 t_{ij}+{x^3\over
1-x}t_{ij}=t_{ij},\quad{\rm for}\ i<j-1
\end{equation} which establishes that
\begin{equation} t_{ij} = x^{i-j+2},\quad{\rm
for}\ i<j-1.
\end{equation}
 [For the special case that $i=j-1$,  we have
from equation  (6.2) that
$2+x=t_{j-1,j}$.]
 This, now, is functionally equivalent to the
similitude relationship assumed by Tokunaga,
namely
\begin{equation} t_{ij} = t_{j-i} =a x^{i-j} .
\end{equation}
 Importantly, the behavior that Tokunaga {\it
assumed}\/ to be valid emerges in a completely
natural way from the underlying mathematics of
our inverse cascade.
 Since $x=0.55495813$, we have for our inverse
cascade
\begin{equation} t_{ij} = \left( 0.55495813
\right)^{i-j+1} \quad.
\end{equation}
 For the deterministic example given in Fig.\ 1,
we have $t_{ij} =
\left( 1/2 \right)^{i-j+1}$.

 Finally, the connection between our treatment
of branching and our earlier treatment of
clustering needs to be established.
 In particular, we observe that $m_j$ turns out
to be equivalent to $n_{1j}$ and that both scale
as $c^{j-1}$ where, as we have already seen, $c
= 1/x^2$.

\section{Adaptation for Fractal Perimeter:
Branching Numbers}

 The branching analysis given in the previous
section is easily modified to include the
fractal perimeter dependence introduced in
equation (4.1).
 Introducing this relation into equations
(5.8)--(5.10), we obtain
\begin{eqnarray}
\dot N_{kj}&=&2 L_{j-1}^2
n_{k,j-1}+\sum_{i=k}^{j-1}\epsilon^{i-j}L_i L_j
n_{ki} \nonumber\\&-&2 L_j^2
n_{kj}-\sum_{i=j+1}^\infty
\epsilon^{j-i}L_i L_j n_{kj},\quad{\rm for}\ k<j,
\end{eqnarray}
\begin{eqnarray}
\dot T_{j-1,j}&=&2
L_{j-1}^2+\epsilon^{-1}L_{j-1} L_j-2 L_j^2
t_{j-1,j}\nonumber\\&-&\sum_{k=j+1}^\infty
\epsilon^{j-k}L_k L_j t_{j-1,j},
\quad{\rm for}\ j>1,
\end{eqnarray}
\begin{eqnarray}
\dot T_{ij}&=&\epsilon^{i-j}L_i L_j-2 L_j^2
t_{ij}\nonumber\\&-&\sum_{k=j+1}^\infty
\epsilon^{j-k}L_k L_j t_{ij},\quad{\rm for}\
i<j-1.
\end{eqnarray} In the steady state, we obtain
analogs of equations (6.1)--(6.3)
\begin{eqnarray} 2 L_{j-1}^2
n_{k,j-1}&+&\sum_{i=k}^{j-1}\epsilon^{i-j}L_i
L_j n_{ki} =\nonumber\\2 L_j^2
n_{kj}&+&\sum_{i=j+1}^\infty
\epsilon^{j-i}L_i L_j n_{kj},\quad{\rm for}\ k<j.
\end{eqnarray}
\begin{eqnarray} 2
L_{j-1}^2+\epsilon^{-1}L_{j-1} L_j&=&
\nonumber\\2 L_j^2
t_{j-1,j}&+&\sum_{k=j+1}^\infty
\epsilon^{j-k}L_k L_j t_{j-1,j},
\quad{\rm for}\ j>1.
\end{eqnarray}
\begin{eqnarray}
\epsilon^{i-j}L_i L_j&=&\nonumber\\2 L_j^2 t_{ij}
 &+&\sum_{k=j+1}^\infty
\epsilon^{j-k}L_k L_j t_{ij},\quad{\rm for}\
i<j-1.
\end{eqnarray}
 Substituting equation (3.4) into (7.4) and
assuming that $x^{j-k}n_{kj}=z^{j-k}$, we
obtain, due to (4.8), an analog of equation
(6.6), namely
\begin{equation} 2xz^{j-k-1}+x^2
\epsilon^{k-j}\cdot{(\epsilon z)^{j-k}-1\over
\epsilon z-1}=z^{j-k}.
\end{equation} Assuming $\epsilon z>1$, we
approximate
$(\epsilon z)^{j-k} -1$ by
$(\epsilon z)^{j-k}$ in the asymptotic limit
$j\gg k$. In this case, equation (7.7)
approximately implies that
$2x+x^2 z/(\epsilon z-1)=z$, or
\begin{equation}\epsilon z^2-(x^2+2\epsilon x+1)
z+2x=0.
\end{equation}
 This latter equation is {\it identical}\/ to
equation (4.10), and has a solution
$z=1/x$, for any $\epsilon$.
 Accordingly, for
$j\gg k$, we have asymptotic self-similarity
with $n_{kj}\approx\beta x^{k-j} z^{j-k}=\beta
c^{j-k}$, where
$c=1/x^2$ as before.

 Substituting equation  (3.4) into equation
(7.6) and dividing by $a^2 x^{2j-4}$, we obtain
\begin{equation}
\epsilon^{i-j}x^{i-j+2}=2x^2 t_{ij}+{x^3\over
\epsilon-x}t_{ij}=t_{ij},\quad{\rm for}\ i<j-1
\end{equation} which establishes that
\begin{equation} t_{ij} =
\epsilon^{i-j}x^{i-j+2}\quad{\rm for}\ i<j-1.
\end{equation}
 [For the special case that $i=j-1$,  we have
from equation  (7.5) that
$2+x/\epsilon=t_{j-1,j}$.]
 Thus, our modification of the Euclidean model to
accommodate fractal perimetric behavior is
complete, and the self-similar description of
the branching process has been shown to follow
in a completely analogous way.

\section{Conclusions and Discussion}

 In this paper, we have presented an inverse
cascade model for clustering.
 This model requires:
\begin{enumerate}
\item The addition of single elements at a
prescribed small scale;
\item The consideration of the clustering
process as a hierarchical tree with side
branching;
\item The probability that a cluster of one
order will coalesce with another cluster of the
same or different order is proportional to the
product of the number of trees of the two orders
and the square root of their masses (or areas);
and
\item Clusters are lost (destroyed) at a
prescribed large scale.
\end{enumerate}

 Our inverse cascade model provides a general
explanation for the behavior of several models
that have been considered to exhibit behavior
which has often been described as
``self-organized criticality'' and occurs in
various settings including the ``forest-fire''
model.
 In this model, the planting of individual trees
is the introduction of single elements, and
coalescence occurs when a planted tree bridges
the gap between two existing clusters.
 The model ``fires'' burn significant numbers of
trees only in the largest clusters and this
terminates the inverse cascade.
 Our model gives the number-mass (or area)
distribution to be $N
\propto 1/A$; this is also found to be the case
for the forest-fire model.
 Our model is also applicable for the sandpile
and slider-block models.
 In the sandpile model, the cluster is the
region over which an avalanche will spread once
it is initiated.
 In the slider-block model, the cluster is the
region over which a slip event will spread once
it is initiated.
 The initiation of an avalanche in the
sandpile model and the initiation of a slip
event in the slider-block model are equivalent
to a spark being dropped on a tree.
 In both models the clusters grow by coalescence.

 We conclude that these models, which are said
to exhibit self-organized criticality, are
neither critical nor self-organized.
 Instead, their behavior is associated with an
inverse cascade which asymptotically approaches
(so long as the largest scales are not involved)
power-law (``fractal'') scaling.
 This behavior is related to the self-similar
direct cascade associated with the
inertial-range of fully-developed isotropic
turbulence.
 This behavior qualifies as a form of
``intermediate asymptotics''\cite{ZB}.
 It is interesting to note that
earthquakes\cite{T99}, landslides\cite{PMBT97},
and actual forest fires\cite{MMT98} also have
analogous power-law frequency-area distributions.

 We have quantified our inverse cascade in terms
of a branching tree hierarchy with side
branching.
 We have adapted the taxonomy used for river
networks to the growth of our clusters.
 The order of each cluster is specified and, in
our mean-field approximation, the number of
clusters of each order is obtained.
 We find that this distribution is identical to
the self-similar side branching distribution
introduced empirically by Tokunaga.
 This distribution has been found to be
applicable for river networks\cite{P95}, DLA
clusters\cite{O92}, and vein structures of
leaves\cite{TPN98}.

\acknowledgments

 We wish to acknowledge the support of NSF Grant
EAR 9804859.
 We are also grateful to Gleb Morein for several
useful discussions.

\begin{figure}
\epsfxsize=\hsize
\centerline{\epsfbox{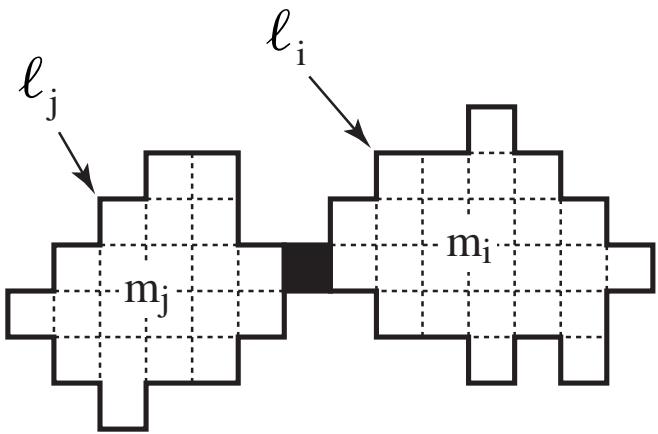}}
\smallskip
\caption{Illustration of how two clusters of
mass $m_i$ and $m_j$ coalesce to form a single
cluster, when an element (solid square) bridges
the gap between two clusters.  
 The two clusters have perimeters $\ell_i$ and
$\ell_j$.
 This example employs a Cartesian lattice for
clarity, although our model does {\it not}\/
require a lattice structure.
\label{fig1}}
\end{figure}

\begin{figure}
\epsfxsize=\hsize
\centerline{\epsfbox{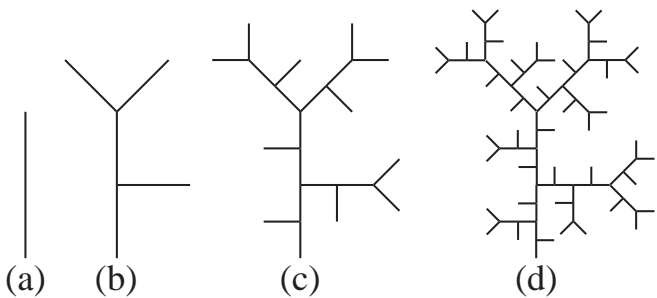}}
\smallskip
\caption{Illustration of the concept of branching
applied to the coalescence of clusters.  
 (a) A single element.  
 (b) Two single elements have been linked to
form a cluster or rank ``2,'' a third element has
joined this cluster as a side branch. 
 (c)  Two clusters of rank ``2'' have coalesced
to form a cluster of rank ``3.''  
 Another cluster of rank ``2'' and two single
elements have been added to this cluster.
 (d) Two clusters of rank ``3'' have coalesced to
form a cluster of rank ``4.''
 Another cluster of rank ``3,'' two clusters of
rank ``2'' and four single elements have been
added to this cluster.}
\label{fig2}
\end{figure}


\begin{references}

\bibitem{BCT92}P.\ Bak, K.\ Chen, and C.\ Tang,
Phys.\ Lett.\ {\bf A147}, 297, B. (1992);\
Drossel and F.\ Schwabl, Phys.\ Rev.\ Lett.\
{\bf 69}, 1629 (1992).

\bibitem{BTW88}P.\ Bak, C.\ Tang, and K.\
Wiesenfeld, Phys.\ Rev.\ {\bf A38}, 364 (1988).

\bibitem{TMMN99}D.L.\ Turcotte, B.D.\ Malamud,
G.\ Morein, and W.I.\ Newman, Physica A {\bf
xx}, yy (1999).

\bibitem{Bak93}M. Paczuski and P. Bak, Phys.
Rev. {\bf E48}, R3214 (1993).

\bibitem{CL89}J.M.\ Carlson and J.S.\ Langer,
Phys.\ Rev.\ {\bf A406}, 470. (1989).

\bibitem{PE99}J.K.\ Parrish and L.\ 
Edelstein-Keshet, Science {\bf 284}, 99 (1999).

\bibitem{BDF99}E.\ Bonabeau, L.\ Dagorn, and P.\
Freon, Proc.\ Natl.\ Acad.\ Sci.\ USA {\bf 96},
4472 (1999).

\bibitem{S57}A.N.\ Strahler, Trans.\ Am.\ 
Geophys.\ Un. {\bf 38}, 913 (1957).

\bibitem{NNN}W.I.\ Newman and L.\ Knopoff, Int.\
J.\ Fracture {\bf 43}, 19 (1990); W.I.\ Newman
and D.L.\ Turcotte, Geophys J.\ {\bf 100}, 433
(1990);  W.I.\ Newman and I.\ Wasserman,
Astrophys.\ J.\ {\bf 354}, 411 (1990).

\bibitem{ZB}G.I.\ Barenblatt and Ya.B.\
Zel'dovich, Russ.\ Math.\ Surv.\ {\bf 26}, 45
(1971); G.I.\ Barenblatt and Ya.B.\ Zel'dovich,
Ann.\ Rev.\ Fluid Mech.\ {\bf 4}, 285 (1972); 
G.I.\ Barenblatt, {\it Scaling, Self-Similarity,
and Intermediate Asymptotics}, (Cambridge, UK:
Cambridge Univ.\ Press, 1996).

\bibitem{H45}R.E.\ Horton, Geol.\ Soc.\  Am.\
Bull.\ {\bf 56}, 275 (1945).

\bibitem{T78}E.\ Tokunaga, Geog.\ Rep.\  Tokyo
Metro.\ Univ.\ {\bf 13}, 1 (1978).

\bibitem{T99}D.L.\ Turcotte, Phys.\ Earth
Planet.\ Int.\ {\bf 111}, 275 (1999).

\bibitem{PMBT97}J.D.\ Pelletier, B.D.\  Malamud,
T.\ Blodgett, and D.L.\ Turcotte, Eng.\ Geol.\
{\bf 48}, 255 (1997).

\bibitem{MMT98}B.D.\ Malamud, G.\ Morein, and
D.L.\ Turcotte, Science {\bf 281}, 1840 (1998).

\bibitem{P95}J.D.\ Peckham, Water Resour.\ Res.\
{\bf 31}, 1023 (1995).

\bibitem{O92}P.\ Ossadnik, Phys.\ Rev.\ {\bf
A45}, 1058 (1992).

\bibitem{TPN98}D.L.\ Turcotte, J.D.\ Pelletier,
and W.I.\ Newman, J.\ Theor.\ Biol.\ {\bf 193},
577 (1998).

\end{references}
\end{document}